September 3, 2018

# The challenges of purely mechanistic models in biology and the minimum need for a 'mechanism-plus-X' framework [1]

**Sepehr Ehsani** [a,b]

[a] *Department of Philosophy, University College London, London, United Kingdom*
[b] *Ronin Institute for Independent Scholarship, Montclair, New Jersey, United States*
ehsani@csail.mit.edu / ehsani@uclmail.net

Ever since the advent of molecular biology in the 1970s, mechanical models have become the dogma in the field, where a "true" understanding of any subject is equated to a mechanistic description. This has been to the detriment of the biomedical sciences, where, barring some exceptions, notable new feats of understanding have arguably not been achieved in normal and disease biology, including neurodegenerative disease and cancer pathobiology. I argue for a "mechanism-plus-X" paradigm, where mainstay elements of mechanistic models such as hierarchy and correlation are combined with nomological principles such as general operative rules and generative principles. Depending on the question at hand and the nature of the inquiry, X could range from proven physical laws to speculative biological generalisations, such as the notional principle of cellular synchrony. I argue that the "mechanism-plus-X" approach should ultimately aim to move biological inquiries out of the deadlock of oft-encountered mechanistic pitfalls and reposition biology to its former capacity of illuminating fundamental truths about the world.

With the advent of molecular biology several decades ago, the gold-standard of explanation in the life sciences has fallen under the umbrella of what are called 'mechanistic' accounts of the molecular pathways underlying a given biological phenomenon being studied. Perhaps surprisingly, while the limitations and challenges that such explanatory models pose "on the ground" in various laboratory-based or clinical experiments are apparent to all, few biologists question the utility and nature of mechanistic accounts and 'models'.[2] Moreover, biologists, and we in general, cannot entertain alternative modes of explanation that depart too much from mechanistic models that are so attuned to our common-sense picture of how the world works.

[1] This thesis was submitted as part of the requirements of the MA programme of the UCL Department of Philosophy. Thanks are due to Dr Luke Fenton-Glynn for his supervision of the project and many helpful discussions. I would also like to thank Adam Rostowski for brining to my attention fruitful literature on situated cognition and for our continued interesting discussions. This work has taken inspiration from a number of papers, specifically Noam Chomsky's "The Mysteries of Nature: How Deeply Hidden" (Chomsky 2009) and "Science, Mind, and Limits of Understanding" (Chomsky 2014), and also from Carl Woese's "A New Biology for a New Century" (Woese 2004).

Meanwhile, philosophers holding views that have given rise to the labels of 'traditional' and 'new' mechanists in the philosophy of biology have engaged with mechanistic accounts in biology mostly at an arm's length, where their analyses (those which are not merely descriptive) have not been borne out by changes of practice or new facets of understanding in the field, or by the ever-present challenges stemming from current mainstream paradigms in biological research.[3] While this might be due to the insular nature of certain areas of science or the translatability of some philosophical frameworks, the time is ripe to present a serious *critique of the philosophical foundation of mechanisms in biology with the view of having a tractable effect on the ways in which biological understanding is acquired and interpreted*.

I have aimed to use arguments that are (i) informed by the history of mechanical philosophy and recent biomedical research, and (ii) modelled on developments in the philosophy of language, cognitive science and generative grammar over the past half-century. This is *not* a survey paper. It makes a very specific point and recommendation in as broad a context as pertinent and achievable. I will argue that the current framework of mechanistic models is tenuous for reaching a broader understanding of the biology of cells under normal and disease conditions, and that an expanded framework that is cognizant of and includes non-mechanistic modes of explanation (e.g., general rules of metabolic scaling or cellular timekeeping) would come closer to the limits to which we can expect to understand and model biological phenomena. Focusing on non-mechanistic accounts of explanation is nothing new in philosophy nor even in biological research, but it is new in the context of modern biological research of the past half-century, and there is much work to be done to reintegrate and develop non-mechanistic elements into current biological research frameworks.

---

[2] A term which will be deciphered in the context of this manuscript in subsequent sections. For a more in-depth discussion, see (Frigg and Hartmann 2018).
[3] There are, of course, exceptions to this generalisation, and some of those exceptions will be shone light on in the relevant sections herein.

## A. WHAT WERE THE MECHANISTIC MODELS IN POST-GALILEAN SCIENCE?

Before delving into the workings of mechanistic models, it is necessary to first say why such models are posited in the first place and why modelling is even important in science. These questions may perhaps seem pedantic to some, but their import and present-day relevance becomes apparent when we take a closer look at the history of scientific inquiry before and after the Enlightenment.[4] The primary occupation of a scientist, and in our case a biomedically-oriented investigator, was set out well by the sixteenth-century physician Paracelsus (1493-1541) who found it imperative for doctors to "inquire of the world" (Lister 1957). Wilhelm von Humboldt (1767-1835) observed in 1792 that "To inquire and to create;—these are the grand centres around which all human pursuits revolve, or at least to these objects do they all more or less directly refer".[5] The aim of such inquiries was, for many centuries, to *directly* understand the natural world. Following the implications of Isaac Newton's (1643-1727) work, the goal of science became—and remains—much more restricted: "the intelligibility of theories *about* the world [and thus understanding the world through the lens of theories]" (Chomsky 2014).[6] Noam Chomsky notes that this transition was "a step of considerable significance in the history of human thought and inquiry, more so than is generally recognized, though it has been understood by historians of science" (Chomsky 2014). This shift to the conceivability of theories of the world (e.g., see (Blumenthal and Ladyman 2018) for an analysis in the history of chemistry) was I think one of the few true "scientific revolutions", to borrow the phrase from Thomas Kuhn, alongside the Galilean approach to science that encouraged puzzlement about otherwise common-sense phenomena. Not many (or any) such paradigm shifts have taken place in science since.

---

[4] I will not discuss the substantial biological contributions of, among others, Aristotle (384-322 BC), Nasir al-Din Tusi (1201-1274) or Baruch Spinoza (1632-1677), for they are beyond the narrow focus on mechanisms as discussed here.
[5] Based on a translation by Joseph Coulthard (1854).
[6] There is a potentially informative analogy between the notion of the "intelligibility of theories about the world" and the German-language concept of *Erkenntnisinteresse*, which is translated as "the scientific ambition of obtaining knowledge about the world" in some sources.

Now let us examine the historical context and the resultant changes of this transition in more detail. Early modern science after Galileo Galilei (1564-1642) held a picture of the natural world that was (and still is) intuitive and common-sensical: that the world is a machine, and that the world is directly understandable. The concept of the universe as a machine was appealing to early Enlightenment scientists, one could speculate, mainly because it is a natural outgrowth of our inherent conception of causation as requiring sequential contact of contiguous parts of an encompassing whole, what David Hume (1711-1776) so ably formalised in his account of causality.

Although much has been done and added with regards to various complementary and divergent theories of causation over the past century, I think that the Humean framework is the most one can hope to achieve in terms of formalising our common-sense *notions and expectations* of causality. One consequence of these notions and expectations is our inherent desire to at least partly view the universe and everything in it as a precise clockwork, composed of metaphorical cogs and wheels. The resilience of this view through time is not only due to its proximity to our common-sense perspective of nature, but also because it implies that humans can take apart the colloquial cogs and wheels and put them back together. The twentieth-century physicist Richard Feynman (1918-1988) famously remarked that "what I cannot create, I do not understand", a dictum that is prevalent thinking in biology, especially protein and synthetic biology (Lupas 2014).

A 'mechanism', stemming from the Greek *mēkhos* ("means"),[7] allows for a *means toward manipulability* of the cogs and wheels, and a means to connect the observed or otherwise perceived inputs and outputs of the system. Laura Franklin-Hall, in her critique of the approach of "new mechanists" in biology, notes that even though "all mechanistic explanations bridge the inputs and outputs of a system with a veridical mechanistic model, there are multiple ways of decomposing a system into organized parts, and multiple mechanistic models, those reflecting different decompositions, can bridge inputs and outputs as required" (Franklin-Hall 2011). This adds an extra layer to the historical appeal of mechanisms other than manipulability, namely, a layer of *choice*. An investigator can choose from a set of possible mechanisms and can therefore experiment amongst a set of possible frameworks. It thus allows room for progress in the explanatory accounts of a scientific inquiry.

---

[7] Etymologies are from *Encarta*, Microsoft Corporation (2009).

The appeal to a generalisable clockwork does not necessarily mean that sixteenth- and seventeenth-century scientists made any rigid claims about the final cause that gave 'life' to the clockwork, i.e., the 'ghost' in the machine. And yet this final cause continued to be a source of great puzzlement. The single most stellar example of the post-Galilean mechanical philosophy is arguably the 'Digesting Duck' (1739) of Jacques de Vaucanson (1709-1782), among his many other automata (**Figure 1**, Appendix).[8] What the duck lacked in a 'living soul' it made up for in having an uncanny resemblance to a physiological creature. Nevertheless, putting the artistic and engineering merits of the great creation of the skilled artisan aside, everyone could see through the appearance and realise the limits of the mechanical duck model. But this was already a significant achievement, for the model informed the viewers of the known facets of a physiological process and concomitantly demonstrated the limits of those known facets.

Overall, the post-Galilean and pre-Newtonian paradigm was this: the part of nature that could be observed, that could be experimented on, and that could be understood can be thought of in terms of mechanistic models. But Newton, *to his own dismay*, undermined this paradigm with his work on gravitation and *action at a distance*. This is an important consideration and should be emphasised. As has been noted by Enlightenment and contemporary commentators alike, perhaps the most significant post-Galilean development in modern science, which was arguably the one true "paradigm shift", was Newton's unintentional undermining of the mechanical philosophy by demonstrating action at a distance and appeals to such "occult" Scholastic notions as gravitational forces and fields:

> "The properties of matter, Newton showed, escape the bounds of the mechanical philosophy. To account for them it is necessary to resort to interaction without contact. Not surprisingly, Newton was condemned by the great physicists of the day for invoking the despised occult properties of the neo-scholastics. Newton largely agreed." (Chomsky 2014)

How could two masses form shared fields of "energy" between them that would allow for attraction and repulsion? What could such fields be made of? These questions, although investigated under various programmes of study to this day, are still very much open questions. In his day, one of the philosophers critiquing Newton was Gottfried Wilhelm Leibniz (1646-1716), who "argued that Newton was reintroducing occult ideas similar to the sympathies and antipathies of the much-ridiculed scholastic science, and was offering no *physical* explanations for phenomena of the material world" (Chomsky 2009). We can perhaps contend that Newton, to borrow the Rylean analogy, "exorcised the machine, leaving the ghost intact" (Chomsky 2014). Chomsky adds that all that we more or less understand (or *can* understand) is usually grouped under the category *physical*, and all else is termed *non-physical* or the *mind*.

---

[8] Mechanical models and constructions follow a long tradition dating back at least to Daedalus, Hero of Alexandria, and others (Koetsier 2001).

Newton's new paradigm, one that "transcended" mechanical philosophy, was taken up by some scientists in the eighteenth century, among them the chemist Joseph Black (1728-1799). He "recommended that 'chemical affinity be received as a first principle, which we cannot explain any more than Newton could explain gravitation, and let us defer accounting for the laws of affinity, till we have established such a body of doctrine as he has established concerning the laws of gravitation'" (Chomsky 2014). Although it took a few centuries for the implications of Newton's achievements to fully permeate into working and hands-on scientific frameworks, the now-almost-unified fields of physics and chemistry eventually began to move away from mechanical models starting in the early twentieth century. The trend in biology, however, has been the opposite. The microbiologist Carl Woese (1928-2012), whose work included studying the Archaean domain of life, and who was critical of the mechanical focus of molecular biology, emphasised the following remark by the theoretical physicist David Bohm (Woese 2004) who observed:

> "It does seem odd […] that just when physics is […] moving away from mechanism, biology and psychology are moving closer to it. If this trend continues […] scientists will be regarding living and intelligent beings as mechanical, while they suppose that inanimate matter is too complex and subtle to fit into the limited categories of mechanism." (Bohm 2009, 34)

Bohm, however, optimistically noted that "in the long run, such a point of view cannot stand up to critical analysis [for] since DNA and other molecules studied by the biologist are constituted of electrons, protons, neutrons, etc., it follows that they too are capable of behaving in a far more complex and subtle way than can be described in terms of the mechanical concepts" (Bohm 2009, 34).

Today, skipping a few centuries from the time of Joseph Black, it is a matter of great puzzlement and concern that these lessons post-Newton have not for the most part continued into our contemporary science, and that there has been a complete reversion to the common-sense notions of mechanical philosophy in biology. Taking a recent example from cancer biology, we read: "Until the 1980s, the reports on the [concept of the tumour microenvironment (TME)] were largely descriptive with little insight into mechanistic issues" (Maman and Witz 2018). The mechanistic issues mentioned simply mean the network cascade of molecules (proteins, RNAs, small molecules, etc.) that form signalling, activation and other pathways within and outside the cell. These are the same cogs and wheels of the clockwork. Looking at another instance, the *Human Cell Atlas Project*, we read that it is an "international collaborative effort that aims to define all human cell types in terms of distinctive molecular profiles (such as gene expression profiles) and to connect this information with classical cellular descriptions (such as location and morphology)" (Regev et al. 2017). The more detailed such an effort is, the more widespread the categories of molecular profiles would be, such that one would eventually have every single cell occupy its own category of specific molecular hallmarks—a trend seen particularly in research on breast cancer (Sparano et al. 2018; Radvanyi 2018). In other words, finding common-denominators based on molecular cogs and wheels would probably not lead to greater understanding, and the right *granularity* and *difference-making properties* would be missed.

Examples of exclusive reliance on mechanistic accounts can be found in all areas of biology: modelling of the blood-brain barrier/neurovascular unit in the central nervous system (Hawkins and Davis 2005) using microfluidic organ chips (Maoz et al. 2018), elucidating mechanisms of T-cell immunological memory (Borghans and Ribeiro 2017), deciphering mechanisms for "memory storage in neurons involving magnetite and prions" (Alfsen et al. 2018), and studying neural mechanisms in general that entail multivariate or network analyses (Giusti, Ghrist, and Bassett 2016). Moreover, statistical and numerical approaches are now becoming much more common in biology, especially computational biology. And, notwithstanding the arguments this paper makes, these methods are even problematic on their own terms: Parsons and colleagues observe that "many studies in the biomedical research literature report analyses that fail to recognise important data dependencies from multilevel or complex experimental designs [and] statistical inferences resulting from such analyses are unlikely to be valid and are often potentially highly misleading" (Parsons, Teare, and Sitch 2018). In addition to the advantages offered by statistical methods, there is a second reason for the increasing reliance on mechanisms in biology: namely, the recent development of molecular tools such as "CRISPR-Cas9"[9] (see (Amoasii et al. 2018) for an example of use in muscular dystrophy research) that have led to a greater *sense of manipulability* of the putative cellular cogs and wheels.

**B.** WHAT IS THE PROBLEM WITH MECHANISTIC MODELS IN BIOLOGY?

To delve into potential hindrances that follow from mechanistic models, it is more important to first ask what the input and 'working materials' of mechanistic models might be and to start from the drawing board. To understand biological phenomena, an investigator typically uses laboratory methods that are, in essence, characterised by culturing cells and all the ensuing techniques.[10] Most observers are familiar with the general framework of laboratory science, where a hypothesis-driven (or perhaps a hypothesis-free) question is investigated using already-established protocols and 'assays' combined with elements of trial and error. But relying on empirical laboratory methods need not always be the case. Theoretical biologists of the first half of the twentieth-century might have generally started with and expanded on thought experiments, in the same spirit that Galileo used thought experiments for most of his scientific achievements. This latter method has a well-established philosophical grounding, for example in Plato (c. 428-348 BC): Clark Glymour and colleagues provide a reminder that *Meno* is "the source of a [philosophical] method: conjecture an analysis, seek intuitive counterexamples, reformulate the conjecture to cover the intuitive examples of the concept and to exclude the intuitive non-examples; repeat if necessary" (Glymour et al. 2010). This, I think, is a perfect recipe for thought experiments.

From these methods, whether empirical or theoretical, we receive *feedback*, *data* or *outputs* that are *ordered* and *hierarchic* in nature. For example, we might observe, or conjure up, that changing variable *A* produces a resulting change in variable *B*, and then a resulting change in variable *C*. But would these changes, if measured using an instrument and perceived via the sensory organs, not be artefacts of the instrument and those senses—which are imprecise and limited? Correspondingly, would realisations from thought experiments not be limited by our inherent cognitive and imaginative limits? This, of course, leads to the instrumentalism versus realism debate, but one that I will not focus on in detail here. In short, however, scientific realism "contends that claims about certain unobservable 'items' which emerge from the theoretical or experimental activity of scientists are literally true" (Carrier 1993). But is the distinction really necessary? Or is 'literally true' just honorific? In this regard Chomsky notes that "the term physical is just kind of like an honorific word, […] like the word 'real' when we say 'the real truth.' It doesn't add anything, it just says 'this is serious truth.' So to say that something is 'physical' today just means 'you have got to take this seriously'".[11] Simon Blackburn makes a similar point, noting that "a difficult question […] is to tell what distinguishes acceptance in a purely instrumentalist spirit from true belief." But pragmatists would deny such a distinction because "in such a philosophy all belief is simply acceptance into the system deemed most useful" (Blackburn 2008).

---

[9] Ignoring for now their potential "pathogenic consequences" (Kosicki, Tomberg, and Bradley 2018).
[10] An informative correspondence regarding an article in the journal *eLife* is relevant in this context: (Chang et al. 2018).

The instrumentalist attaches more truth-conditions to the explanandum, noting the scope and limits of the measuring instruments, the perceiving human sense(s) and our inherent cognitive capacities. We can never know if our explanation is the *real* truth, or if there are more truths hidden beneath the surface yet to be discovered. We can hypothesize various explanatory frameworks that would fit our data and conditions, and, as is now more popularly discussed, *infer to the best explanation* (to be discussed in subsequent sections).

Let us now take a more concrete example. If we are performing a cell-based colourimetric assay that requires the use of a spectrophotometer to take readings, surely the data that we receive are limited by the constraints of the instrument, the spectrum that it can read through and measurement errors. Therefore, is this not a problem? And are we not moving away from the *reality* that lies in the cells being used for the assay? I think the answer here is negative, because

---

[11] "On the Philosophy of Mind," interview at State University of New York at Stony Brook (2003).

Page 9

biology, just like physics, is not a "metre-reading science".[12] The experimenter must always be cognizant of the limits of the instrumentation and the limits of the question being posed. Overall, although there is much more to say about this topic, I do not think there is a dichotomy of instrumentalism versus realism when it comes to our analysis of mechanisms and models in biology.

Having received feedback or data from the laboratory-based or thought experiment, we can then proceed to pose a theory, a *collection* of which can then form a comprehensive theoretical *model*, which could either be mechanistic or non-mechanistic. I will return to non-mechanistic theories at a later stage, but for now the focus will be on mechanistic theories. Here we should note that although biologists using the term "mechanistic" may not necessarily be familiar with its connections to Cartesian and Pre-Newtonian science, mechanistic theories, as we will see, suffer from some of the same problems that late-Enlightenment philosophers aimed to alleviate. William Bechtel provides a current definition of a mechanistic theory as follows:

> "A mechanism is a structure *performing a function* in virtue of its *component parts*, component operations, and their organization. The orchestrated functioning of the mechanism is responsible for one or more phenomena […] Mechanisms *exist in nature*, whereas mechanistic explanation involves an investigator presenting an account of the mechanism taken to be responsible for a given phenomenon." (Bechtel 2008, 160)[13]

This, to me, sounds like an uncontroversial description, one that emphasises a collection of *components* performing one or more *functions* which together represent and give hints to an *actual mechanism* that exists in nature. Understanding by manipulating the components of a system and receiving feedback is common practice in all areas of biology (e.g., (Krug, Salzman, and Waddell 2015)).

What could be some problematic facets of the above description of mechanisms? One is the notion of *function*, which implies teleology, in biological phenomena that we are only beginning to superficially understand. I will return to this later. The second is the assumption that a theoretical framework imposed by our cognitive powers (e.g., the concept of 'mechanism') somehow exists

---

[12] For an analysis of the move in biology toward a "data-driven" science, see Sabina Leonelli's *Data-Centric Biology: A Philosophical Study* (University of Chicago Press, 2016). The phrase "data-driven" or "data-centric" is problematic, for if we take "data" to be synonymous with pieces of evidence, the assumption has always been that *evidence* is at the centre of any serious attempt at scientific inquiry.

[13] Emphases in quotations are mine, not the author's, unless otherwise indicated.

*ipso facto* in nature, again hinting at the debate on realism. I remain sceptical that the *ipso facto* existence of mechanisms or any other framework can be the case, for we just can never "really know for sure" (and beyond what we are capable of knowing). This latter part of the sentence sounds like (and is) a truism, but one that is often neglected. Saying that mechanisms truly exist in nature is somewhat an 'honorific' statement. For nomological accounts, at least, even though the same 'unknowability' concern applies, we might speculate that nomological rules such as thermodynamic laws are more entrenched in nature than instantiated mechanistic accounts that are so context-dependent. Moreover, going beyond the philosophical basis and looking at the on-the-ground impact of mechanistic thinking in biology, the result is quite mixed. While there are enthusiastic receptions of (and findings from) mechanistic understanding for certain aspects of normal cell biology and pathobiology (see e.g., instances from research on paediatric leukaemia (Greaves 2018)[14] and gene therapy for β-thalassemia patients (Thompson et al. 2018)), in many other respects the outcome has been disappointing (for a more biologically-oriented discussion, see e.g., (Ehsani 2016) and (Ehsani 2018)).[15] In 2004, Woese aptly summarised the challenge of mechanistic biology as follows:

> "The machine metaphor certainly provides insights, but these come at the price of overlooking much of what biology is. Machines are not made of parts that continually turn over, renew. The organism is. Machines are stable and accurate because they are designed and built to be so. The stability of an organism lies in resilience, the homeostatic capacity to reestablish itself. While a machine is a mere collection of parts, some sort of 'sense of the whole' inheres in the organism, a quality that becomes particularly apparent in phenomena such as

---

[14] This review article presents "evidence supporting a multifactorial causation of childhood acute lymphoblastic leukaemia".

[15] Let us briefly visit three recent examples of these challenges in more detail: (i) One instance is an analysis of the effect of omega-3 fatty acids (for which many mechanistic models in a molecular biological context have been studied), that concluded the following: "Moderate- and high-quality evidence suggests that increasing EPA [eicosapentaenoic acid] and DHA [docosahexaenoic acid] has little or no effect on mortality or cardiovascular health (evidence mainly from supplement trials). Previous suggestions of benefits from EPA and DHA supplements appear to spring from trials with higher risk of bias. Low-quality evidence suggests ALA [alpha-linolenic acid] may slightly reduce CVD [cardiovascular disease] event risk, CHD [coronary heart disease] mortality and arrhythmia" (Abdelhamid et al. 2018). *This example points to problems with the putative mechanisms of cardiovascular disease.* (ii) The second example pertains to antidepressant drugs for the treatment of major depressive disorder: "While this meta-analysis (Cipriani et al. 2018) is a meticulous and comprehensive summary of multiple therapeutic options, the clinical scope is limited and it only confirms (with greater precision) findings of other reviews […] Patients and clinicians need clear answers to many more questions before shared decision making in this area is fully informed" (McCormack and Korownyk 2018). *This case hints at challenges with putative mechanistic models of psychiatric conditions.* (iii) Lastly, we can point to lingering questions that usually persist upon the conclusion of clinical trials: "Cast in terms of a courtroom trial […] the halt of almost any clinical trial […] declares not a verdict but a mistrial" (https://mosaicscience.com/story/deep-brain-stimulation-depression-clinical-trial/). *This example points to problems with clinical trials aimed at understanding the efficacy of therapies for which a mechanism of action is thought to have been well-established.*

regeneration in amphibians and certain invertebrates and in the homeorhesis exhibited by developing embryos." (Woese 2004)[16]

It would be useful to illustrate the above challenges using a running example from current biomedical research. One such candidate might be the ever-more-urgent topic of antibiotic resistance, tolerance and persistence (Meylan, Andrews, and Collins 2018; Brauner et al. 2016). Also, striking examples can be found in cancer biology, where "most new cancer drugs are failing to deliver any clinically meaningful benefit" (Cohen 2017). But I will use examples from the field of Alzheimer's disease (AD) research which can be illustrated in a more focused way. For a number of decades, the dominant hypothesis in AD research has been centred around the misfolding and aggregation of a protein fragment called amyloid-β, a process which is thought to have both genetic and environmental components, and which eventually leads to neuronal death. Other proteins, including one termed Tau, are also thought to be involved in the pathogenesis of AD, but amyloid-β has been the primary focus. Despite many large trials testing this hypothesis, none has so far led to a non-symptomatic therapy that can move in the direction of "curing" the disease, i.e., tackling AD at its root cause(s). Many other avenues have also been tried, including focusing on the protein γ-secretase (Tagami et al. 2017), the modulation of the brain's neurotransmitters (Bennett 2018), groundwater lithium levels (Parker et al. 2018), physical fitness (Lamb et al. 2018) and physical activity (Brasure et al. 2018), and the human herpesvirus (Eimer et al. 2018; Readhead et al. 2018). Published trends in AD clinical trials this year "include *more trials in preclinical and prodromal populations* and greater use of biomarkers to support the diagnosis of AD" and "there is an increase in nonamyloid mechanisms of action for drugs in earlier phases of drug development" (Cummings et al. 2018).[17] There is also now greater attention to alternative hypotheses in the Alzheimer's research field (Cubinkova et al. 2018).

---

[16] Genes are now often thought of as acting as transferrable components of disease mechanisms. See for example (Hui et al. 2018) for an analysis of a gene now thought to play a role in both Crohn's and Parkinson's diseases.

[17] The interpretation of clinical trial results is in itself a challenging task. This very current example is illuminating: "The therapy, developed by Biogen Inc. in Cambridge, Massachusetts, and Eisai Co. Ltd. in Tokyo, is an antibody that binds to and helps clear out a protein fragment called β-amyloid, which builds up in the brain and is thought to drive the disease's neurodegeneration. The statistically complex phase II trial in 856 people with an early form of Alzheimer's failed to show benefit after 12 months by the measure the drug's sponsors selected as the trial's primary endpoint. But after 18 months, the 161 patients getting the highest of five doses had reduced brain amyloid buildup, and their cognitive skills declined 30% more slowly than those of patients getting a placebo, Eisai announced. But only that small group showed a statistically significant benefit, and researchers decided midstudy

Comparing proposed mechanisms of amyloid formation in AD in 1989 and 2015 (**Figure 2**, Appendix) reveals a (some might say expected) truth that while the precision and detail of the model has greatly increased, the general structure of the mechanism has remained remarkably stable. I think something must change, for the sake of the philosophy of biology and, more importantly, for patients' sake. The primary issue, I believe, is that we as humans appear to be inherently disposed to assign mechanisms to natural phenomena, i.e., we are *methodologically mechanistic*. This is similar to our inherent predisposition to see dichotomies such as random/structured or ordered/disordered (even though these may occlude events that lie outside of both categories), or to perceive a duality between mind and body (Phillips, Beretta, and Whitaker 2014), again despite the fact that Newton showed there to be no such concept as a 'body': "If metaphysical dualism has been undermined, what is left is a kind of *methodological dualism*, an illegitimate residue of common sense that should not be allowed to hamper efforts to gain understanding into what kind of creatures we are" (Chomsky 1995). It is also not surprising that a mechanical model might be seen to almost be a true image of the world, reminiscent of aspects of Vaucanson's 'Digesting Duck' alluded to earlier. Interestingly, the poet James Whitcomb Riley (1849-1916) wrote of a 'duck test', now famous, in the early 1900s as follows: "When I see a bird that walks like a duck and swims like a duck and quacks like a duck, I call that bird a duck". What is more, because we are inherently mechanistic, the very fact of assigning any mechanism to a biological phenomenon is quite normal among investigators and is essentially an unfalsifiable act. The utility of mechanisms is taken as a tautology in modern molecular biology. This is unjustified due to the history behind the pitfalls of mechanical philosophy and the many cyclical problems in biomedicine that seem reticent to resolution. The burden of proof must therefore lie with the mechanist to falsify non-mechanistic alternatives, not the other way around.

---

to remove people with an Alzheimer's-promoting gene variant from that group—leading many in the field to await a larger trial to confirm any disease-slowing effects" ("Alzheimer's drug buoys hopes," *Science* 2018, 361(6401):433).

## C. WHAT ARE THE INCONTROVERTIBLE ELEMENTS OF A MECHANISTIC MODEL?

The data or feedback that an investigator receives from their thought or laboratory experiment is, in the first instance, hierarchical and in different *levels*. The notion of levels that is intended here could either be (i) temporal, in the sense that the investigator becomes aware of the effects of one variable *before* the effects of another; or spatial, such as variables that might be active inside and outside the cell. A tertiary sense of *levels* could also be that of scales, where some variables might be present or active at a more 'macro' scale compared to others. The hierarchical organisation of the primate visual system (Arcaro and Livingstone 2017) is one example of a system where there is both temporal and processing-complexity differentiation in cascading levels.

The order in which the feedback from the experimental system is received (or postulated to be received in the case of thought experiments) is prior to forming mechanistic or other theories, and the model then acts as a synthesizer of this level-separated information. Bechtel observes that "although the notion of levels has proved a vexed one […] there is a clear sense in which the parts and operations within a mechanism are situated at a lower level of organization than is the mechanism itself" (Bechtel 2008, 160). Then there is *regularity*, in the sense that changes in one variable are always accompanied by changes in another (i.e., they are *correlated*), or that one set of changes in a set of variables always predictably leads to changes in another set (i.e., positive or negative feedback loops). In an important paper in 2000, Peter Machamer, Lindley Darden and Carl Craver note that "what makes [a mechanism] regular is the *productive continuity* between stages" (Machamer, Darden, and Craver 2000). An overarching theme of their paper is that mechanistic models make regular biological *processes* more intelligible. Hierarchy of data levels and regularity of processes I think should be uncontentious elements of an ideal theoretical model in biology.

There are other elements, however, that are more problematic on closer inspection. One is what we might call 'manual recreatability', reminiscent of the earlier quotation by Feynman. A mechanism, as in a clock, implies that if individual components were placed one by one in the right niche and connected with the other components in exactly the right manner, the phenomenon the mechanistic theory sought to model could be recreated anew. But this is simply not the case.

Synthetic biology (which includes synthesizing protein and gene networks (Clore 2018)), and systems biology (Matthiessen 2017), cannot yet recreate seemingly simple traits in model organisms such as nematodes. This is probably because there are certain aspects of biological phenomena that cannot fundamentally be captured by mechanisms, and also perhaps because there is seemingly indefinite redundancy and control. Bechtel characterizes these circumstances by noting that "there is a complex[18] web of control operative on biological mechanisms [and that the] philosophical analysis of mechanisms needs to be extended from focusing exclusively on how primary mechanisms produce the phenomenon for which they are responsible to how they are controlled" (Bechtel 2018). Interestingly, it is informative to compare these thoughts to Bechtel's earlier characterisation of mechanisms in 2008, when he wrote: "Biological systems are typically bounded,[19] and there are good reasons for the mechanisms within them, including cognitive ones, to be segregated from one another" (Bechtel 2008, 167). Isolating mechanisms as a means of convenience and intelligibility for the investigator might be useful, but as Bechtel points out in the former, more recent, quotation, the reality is that mechanisms as we conceive of them are mind-bogglingly interconnected for any given phenomenon.

In my quest to argue that a Feynman-style mode of understanding (i.e., "what I cannot [re]create, I do not understand") is not possible in the biological sciences, let us look at a relevant example from the burgeoning field of 'machine learning' and computational literary analysis. Patrick Winston, a computer scientist, has aimed to "build computer systems that can simulate the human mind's unique powers of perception and insight".[20] To 'simulate', in this case, is the same as to 'model'. In doing so, Winston is using William Shakespeare's *Macbeth*, and the bare components of the story that he is using in the 'artificial intelligence' project has the following form:

> "Macbeth is a thane and Macduff is a thane. Lady Macbeth is evil and greedy. Duncan is the king, and Macbeth is Duncan's successor. Duncan is an enemy of Cawdor. Macduff is an enemy of Cawdor. Duncan is Macduff's friend. Macbeth defeated Cawdor. Duncan becomes happy because Macbeth defeated Cawdor. […] Macduff curses Macbeth. Macbeth refuses to surrender. Macduff kills Macbeth. The end." (Winston 2016)

---

[18] The word "complex" is usually a signal for "we do not currently understand", and it is thought-provoking how many times this term is used in various biological manuscripts.
[19] All biological systems, including our linguistic faculty which has the property of discrete infinity, are bounded.
[20] Rosen, A. "Toil and trouble: How 'Macbeth' could teach computers to think". *The Boston Globe*, 2 June 2018 (https://www.bostonglobe.com/metro/2018/06/02/toil-and-trouble-how-macbeth-could-teach-computers-think/h9kbBEX7BVkp4Rc6XbJrPM/story.html).

Although Winston "cautions that this short, rough summary 'is meant to facilitate research, not to be a faithful rendering of the nuances of Shakespeare'",[21] can this simulation, which we might metaphorically call 'mechanistic', ever even begin to move in the direction of recreating the nuances of Shakespeare's works, i.e., to recreate what makes them timeless pieces of literary genius?

Another problematic implication of mechanistic models is that of causal contact. As discussed earlier, our Humean common-sense notion of causation implies that two components of a mechanistic model that are contiguous, and one is temporally prior to the other, would in all likelihood be in a causal dyad. But the notion of 'contact' in a mechanism is problematic. At its core, 'contact', a concept borrowed from human language with an intractable semantics (from the Latin *tangere*, meaning "to touch"), is vacuous at best at a molecular level. The meaning of contact and 'interaction' between two proteins or small molecules is merely a matter of convention, for it is not a settled matter how close in distance and for what period of time two molecules would have to be adjacent to each other and under what conditions for an 'interaction' to take place (i.e., one molecule affecting the structure and working(s) of another).[22] Furthermore, when we move from molecular and cell biology to the level of disease aetiology (Chang et al. 2018; Kell and Pretorius 2018) and epidemiology (Poots et al. 2017; Krieger and Davey Smith 2016), the *criteria* and the *candidates* for assigning causality can take on completely different personas:[23] they become matters of definition, convention or utility to the application at hand. We thus lack a universal common denominator in causation and causal proximity as an element in mechanistic models.

Analogous to the discussion on causality, there are also hints of teleology and function as elements of mechanistic theories. Assigning functions to components of a mechanism, such as cogs and wheels, again stems from our common-sense notions of causality and mechanism, and is in line with an Aristotelean view of nature. There is in fact a well-established discipline of assigning structure-function relationships in protein biology (Comstock et al. 2015). But can 'function' stand a thorough philosophical analysis? For example, what is the 'function' of the heart? Is it "merely" to pump blood (Weber 2017)?[24] Or can it also do other things? Ogawa and de Bold, for example, point out that "the discovery of the endocrine heart provided a shift from the classical functional paradigm of the heart that regarded this organ solely as a blood pump to one that regards this organ as self-regulating its workload humorally and that also influences the function of several other organs that control cardiovascular function" (Ogawa and de Bold 2014). Function can only make sense within the mechanism at hand, and its transplantation to the biological phenomena under investigation devoid of the context of the mechanism could prove problematic. Machamer and colleagues also note that functions "should be understood in terms of the activities by virtue of which entities contribute to the workings of a mechanism" (Machamer, Darden, and Craver 2000). This has in my view positive implications for teleology as well, where "working toward an end" would depend on the exact context and the precise mechanistic model under study.

---

[21] "Think you have trouble reading Shakespeare? Here's how an MIT scientist boiled it down for a computer". *The Boston Globe*, 2 June 2018 (https://www.bostonglobe.com/metro/2018/06/02/think-you-have-trouble-reading-shakespeare-here-how-mit-scientist-boiled-down-for-computer/xdAdSaJq8BKfDd241A1JVI/story.html).
[22] An example of a problematic interpretation of contact and interaction pertains to the export of the proteins TDP-43 and FUS (implicated in amyotrophic lateral sclerosis and frontotemporal dementia) from the cell nucleus via passive diffusion versus active transport (Ederle et al. 2018). Do the proteins interact with the nuclear membrane, with certain proteins on the membrane, with both, and if so, what is the effect? Is one interaction more consequential than another? Ederle and colleagues conclude that "both proteins are exported independently of the export receptor CRM1/Exportin-1".
[23] See also (Conley and Zhang 2018).

Nevertheless, there may be metaphysical objections to why function and teleology should not be analysed here with the same brush: "one might (controversially) suggest that our spatial vision is for success at coping with a spatial world, whereas colour vision may not be for success at coping with a coloured world, but adapted to the skilful tracking of surfaces through changes of light, and this would be a way of defending a primary/secondary quality distinction" (Blackburn 2008). In a 1943 paper in *Philosophy of Science* entitled "Behavior, Purpose and Teleology" (an article that spurred the beginnings of the field of 'cybernetics'), the authors concluded that:

> "The concept of teleology shares only one thing with the concept of causality: a *time axis*. But causality implies a one-way, relatively irreversible functional relationship, whereas teleology is concerned with behavior, not with functional relationships." (Rosenblueth, Wiener, and Bigelow 1943)

There is a further note of caution about what we mean by teleology in the adjective. In discussing debates on teleology in biophysics in the 1930s, Phillip Sloan points to the evolutionary biologist Ernst Mayr's distinction between 'teleonomic' and 'teleomatic' explanations. In Mayr's view, Sloan writes, 'teleomatic' accounts entail an endpoint (such as "the cooling of hot iron") but not a goal. On the other hand, a 'teleonomic' process "'owes its goal-directedness to the operation of a program,' [which] implies the interaction of intrinsic properties within the organism with external laws and conditions" (Sloan 2012). This *teleonomic* framework is quite a promising element, and we will return to it in the following section.

---

[24] Marcel Weber rightly points out that "our valuing survival or other goal states may be the reason why biology *seeks* functional knowledge" (Weber 2017).

### D. ADDITIONAL ELEMENTS BASED ON PUZZLEMENTS AND COMMON-SENSE DEPARTURES

Now that we have examined some promising and, on the other hand, certain problematic elements of current mechanistic models in biology, the issue at hand is to propose an alternative framework, which I will be calling a 'mechanism-plus-X' approach. This means a theoretical model that takes the essential elements of mechanistic theories and merges them with a contextually-appropriate non-mechanistic element—call it X. There is nothing mysterious about X, and this is not a reference to incorporeal substances. X is, to put it concisely, a *principle*. I will explain. "A skilled puppeteer can give 'life' to a marionette by pulling on a few strings connected to individual limb joints", write Louis and Simpson in a recent commentary (Louis and Simpson 2018). But the nervous system "faces an analogous challenge: sensory information integrated in the brain has to be conveyed to the motor neurons and muscles through a limited number of descending neurons". One can either inquire about *the nature of the puppeteer* or, as a lower-hanging fruit, ask about *the principles* based on which the puppeteer operates the marionette. The latter is more achievable and is more imminent and pressing. This is what X entails. It is arguable whether these principles are akin to notions such as "activity" and "capacity". I am unclear on this, as such notions carry a diverse body of interpretation. Nonetheless, these notions as described by Alan Chalmers are close to what we could call principles: "unless we attribute some source of activity to matter, that is, some powers or capacities, we cannot hope to account for, or even accommodate, the wealth of activity in the world, such as chemical reactions, the swing of a pendulum or the running of an electric motor" (Chalmers 1993).

How could these principles be discovered, and what is their nature and relation to current mechanistic models in biology? The natural sciences, if done properly, usually start with a puzzlement. For example, one might become puzzled by the diversity of colours in fleshy fruits (Stournaras et al. 2013), or possible modes of non-genetic inheritance (Newby and Lindquist 2013). The first route to study the puzzle is then through common-sense methods, but almost invariably, upon closer inspection of the puzzle, there is a departure from common-sense and, hopefully, the realisation of new insight or understanding.[25] This process, and the allowance to depart from common-sense frameworks in a mechanism is what I hope would be captured by X.

Going back to Alzheimer's disease, a source of puzzlement one could immediately ponder is to ask why it is only humans who appear to have the capacity to develop AD:

> "Like many humans, non-human primates deposit copious misfolded Aβ [amyloid-β] protein in the brain as they age. Nevertheless, the complete behavioral and pathologic phenotype of Alzheimer's disease, including Aβ plaques, neurofibrillary (tau) tangles, and dementia, has not yet been identified in a non-human species. Recent research suggests that the crucial link between Aβ aggregation and tauopathy is somehow disengaged in aged monkeys." (Walker and Jucker 2017)

And I should note that this 'puzzlement' to 'common-sense' to 'departure' outline is not restricted to the natural sciences. The eighteenth-century mathematician Leonhard Euler appears to have followed the same route when proposing a value for the infinite series $1−2+3−4+5−6+… (=1/4)$ or for the infinite sum of natural numbers $(1+2+3+…=−1/12)$ (Kaneko, Kurokawa, and Wakayama 2003), departing from the common-sense notion of summation:

> "[…] when it is said that the sum of this series $1−2+3−4+5−6$ etc. is $1/4$, that must appear paradoxical. For by adding 100 terms of this series, we get $−50$, however, the sum of 101 terms gives $+51$, which is quite different from $1/4$ and becomes still greater when one increases the number of terms. But I have already noticed at a previous time, that it is necessary to give to the word sum a more extended meaning. We understand the sum to be the numerical value, or analytical relationship which is arrived at according to principles of analysis, that generate the same series for which we seek the sum."[26]

One way of encouraging these new insights to result from a new theoretical framework is to move beyond mere deductions from mechanisms toward inference to the best explanation (IBE) (Mackonis 2013) and 'abductive' reasoning, where "induction seeks facts to test a hypothesis [whereas] abduction seeks a hypothesis to account for facts".[27] Such a framework would also be cognizant of oft-forgotten rules in biological research, that for example the "absence of evidence is not evidence of absence" (Altman and Bland 1995).

---

[25] For a more thorough discussion, see (Chomsky 2015).
[26] Translation of 1768 manuscript "Remarques sur un beau rapport entre les series des puissances tant directes que reciproques" by Lucas Willis and Thomas J. Osler (eulerarchive.maa.org/docs/translations/E352.pdf).

An IBE and/or abductive approach is antithetical to the view of biology or any other natural science as a "metre-reading science". Let us take an example: it has been commented that the late biologist John Sulston (1942-2018) "recognised that he was not the kind of scientist who dreamed up hypotheses of potentially explanatory force and put them to the test. Instead he saw his science as principally Baconian, or, more unflatteringly, ignorance-driven: comprehensive collection of data, which then provides a resource for others to test their own hypotheses".[28] But can such a seemingly 'hypothesis-free' approach really exist? I do not think what any scientist does could be 'hypothesis-free' at all. For even a purely systematic task such as sequencing a part of the genome requires at least a number of underlying and tacit hypotheses, such as where to sequence and why, or what to discard as noise and what to include as genuine data. As Chomsky points out, "non-science starts with just collecting data and trying to make inductive generalizations from it, and it gets absolutely nowhere".[29] Regardless of whether hypotheses are explicitly stated at the outset of a scientific investigation, their implied presence should always be assumed.

Let us delve a little deeper into what X could be, which should clearly be a nomological non-mechanistic entity, in the form of invariant generalisations or general operative principles (and subsets thereof, such as generative principles that have been formalised, e.g., in generative grammar). To put this into context, Elliott Sober noted about twenty years ago that "perhaps it is time to investigate the possibility that biology has no empirical laws of evolution because of the strategies of model building that biologists have adopted" (Sober 1997). Additionally, Bechtel and Abrahamsen investigated scenarios where mechanisms generalise without appealing to laws (Bechtel and Abrahamsen 2005). Here I would like to stay clear of linking X to any 'laws' or formalisations, and instead focus on *weaker versions* of 'laws' in the form of operative rules and principles in biology (in theory, such principles could eventually become 'laws' if shown to generalise under various conditions and paradigms). James Woodward, for example, argued that "invariance rather than lawfulness is the key feature that a biological generalization must possess if it is to figure in explanations" (Woodward 2001). Usually, when a 'law' is stated in biology, which happens much less often than in chemistry and physics,[30] the focus becomes tuned to the intricacies of the purported law itself and the sense of puzzlement, as discussed earlier, quickly dissipates. In fact, there are only a few relatively *general* principles in biology—ones that are specific to biology (Koonin 2012): One might be allometric/metabolic scaling (Ballesteros et al. 2018; Barabasi 2017; Gearty, McClain, and Payne 2018)[31] and another could be psychology's "Miller's law", an empirical 'invariance' of the number of items capable of being held in working memory (Miller 1956). In the case of mathematical formalisations (e.g., the 1952 Hodgkin-Huxley model based on differential equations modelling the initiation and propagation of action potentials in neuronal cells), Roman Frigg points out that "one gets so used to the mathematics that one suddenly forgets that there are other elements to a complete model" (Frigg 2002).

---

[27] This wording is from: "Abductive reasoning", *Wikipedia* (https://en.wikipedia.org/wiki/Abductive_reasoning).
[28] Ferry, G. "Sir John Sulston obituary". *The Guardian*, 11 March 2018 (https://www.theguardian.com/science/2018/mar/11/sir-john-sulston-obituary).
[29] "Language and Other Cognitive Systems: What is Special about Language?" Lecture at the University of Cologne (2011).

    I think one of the promising areas that X could be grounded in should be in the form of operative principles based on one of the few uncontroversial 'laws' of nature (recall the 'teleonomic' processes discussed earlier), i.e., the 'law' of *computational efficiency*. This rule is well known at the scale of organisms. For example, we know of the seeming appearance of probability calculations by pigeons: "computation of probability may be common to many classes of animals and may be driven by the need to forage successfully for nutritional food items, mates, and areas with a low density of predators" (Roberts, MacDonald, and Lo 2018). There is also much work on the principles of human brain computation: "Perhaps the brain really does compute in much the same sense as, and in much the same way as, a conventional computer. Why should the conveyance of acquired information proceed by principles fundamentally different from those that govern the conveyance of heritable information?" (Gallistel 2017). But I would like to concentrate on putative computational principles at a cellular level. These might include, for example, computational principles in reaction-diffusion systems as in vertebrate limb development (Newman

---

[30] There is always a 'to and fro' in physics about what laws actually are. Looking at the article "What if there are no laws of nature? How nothingness can explain everything about reality" (*New Scientist*, 11 November 2017), one wonders how there *cannot* be any law (even nothingness). The question is to what extent the laws are ever knowable.
[31] See also (Elgin 2006).

2007), protein folding and amino acid affinity principles (Rackovsky 2009), and the notion of time calculations in the cell (Ehsani 2012). These computational processes would also be bound by and delimited by physical constraints of the cell, for example the fact that cells are round (Wang, Dandekar, et al. 2018).

Computational principles would fit well with the first level in the computational neuroscientist David Marr's (1945-1980) three levels of analysis: computational, representational/algorithmic and implementational (Marr, Ullman, and Poggio 2010).[32] Furthermore, there is a great deal of literature and precedent of similar thematic work in biolinguistics, having to do with generative computational principles: "a generative grammar must have the recursive resources necessary to generate a non-finite set of sentences out of a finite set of lexical items" (Jacob 2016). Going back to the earlier reference to a 'puppeteer', one could ask where such computational principles come from, akin to Nancy Cartwright's position that "It takes […] a *nomological machine* to get a law of nature" (Cartwright 1997). This appears more of a mystery than an approachable problem at present, and in any case, I have little to say on this question.

Situating the 'mechanism-plus-X' framework in a generative context via X would make our model more informative, and more likely to initiate insights on the study question. In the same line, Darden suggests that we need to "situate molecular biological mechanisms, such as DNA replication, into a wider hierarchical context" (Darden 2008). Returning to a case noted in an earlier footnote on omega-3 fatty acids, the cardiovascular biologist Tim Chico reminds the reader that "previous experience has shown that although some types of diet are linked to lower risk of heart disease, when we try to identify the beneficial element of the diet and give it as a supplement it generally has little or no benefit".[33] Context and circumstances, therefore, are everything in biology due to the plethora of known and unknown variables involved, and as a result I think the mantra of *ceteris paribus* can rarely be true in biology as a matter of principle.[34] Other things (i.e., as part of the context) are usually never exactly equal, since unknowns cannot be controlled.

---

[32] See also (Bechtel and Shagrir 2015).
[33] Boseley, S. "Omega-3 no protection against heart attack or strokes, say scientists". *The Guardian*, 18 July 2018 (https://www.theguardian.com/society/2018/jul/18/omega-3-no-protection-against-heart-attack-or-strokes-say-scientists).
[34] For some discussion on the concept of context in cell biology, see (Ehsani 2013).

Let us look at two examples. In the case of the prion protein, which is mainly implicated in mad cow disease in cattle and Creutzfeldt-Jakob disease in humans, but may also be involved in AD, an important question has concerned the seeming diversity of the molecular environment (i.e., cellular context) in which it resides. It now seems that the *context* under which the prion protein's own molecular context is analysed holds the key to the diversity: "the molecular environment of a given protein-of-interest can be surprisingly diverse when comparing distinct [cell line] models" (Ghodrati et al. 2018). A second case in point concerns the neural wiring "map" of nematodes, also known as the "connectome". A recent commentary on research on this topic states that "by pairing the connectome with functional tests—like killing neurons or turning them on and off—researchers are uncovering how neural circuits work".[35] But it continues with the following caution:

> "Overlying the connectome are layers of fluctuating complexity that can't be captured by a static map. The connectome cannot indicate whether neurons are working with or against each other. It does not record changes in synaptic strength, a measure of how much influence neurons have over their neighbors. And on top of all this, there is the rush of neuromodulators."[36]

Lastly, but most importantly, how would this new framework work exactly? Let us consider **Figure 3A** (Appendix) as a starting point. This is a classic and very simplified and uncomplicated mechanism. Now let us take two of the unproblematic elements of mechanisms described earlier, regularity and hierarchy, and remove the arrows (**Figure 3B**). This is a bare-bones structure, where the regularity is demonstrated by the proximity—and hence the *degree of correlation*—of the variables to each other, and hierarchy is demonstrated in the levels and therefore the priorities of the variables in reference to one another. For the sake of argument, the levels in this figure can refer to spatial levels, where we might have intracellular, extracellular and lymphatic vessel spaces as three different physical levels. Here the position of component *D* is noteworthy, as it appears to straddle two levels and necessitate further investigation. Now we can ask how this configuration actually changes (experimentally or theoretically) when X,[37] here set to be a putative cellular timing condition, is at α (e.g., the ambient temperature is very low and the plasma membrane of the cell is more stationary).[38] Then we determine the configuration under X = β, where β, let us assume, is the performance of the cellular pacemaker at room temperature. Then for X = γ, perhaps the cell is still at room temperature but is now in a cultured dish as part of a 'syncytium' and therefore affected by the conjoining plasma membranes of other cells. The way in which one would then superimpose the configurations for different values of X produces a semi-dynamic model that points to the relations and effects of the different variables in a more realistic and perhaps insightful manner, one which could also point to hidden variables that would otherwise escape the investigator's notice.

---

[35] Chen, J. "In lofty quest to map human memories, a scientist journeys deep into the mind of a worm". *STAT*, 13 August 2018 (https://www.statnews.com/2018/08/13/connectome-neurology-mapping-cognition/).
[36] Ibid.
[37] Here setting X at different values is congruous to such usages in mathematics, where, for example, the 1−2+3−4+… infinite series alluded to earlier is a special case of $1 - 2x + 3x^2 - 4x^3 + \ldots$ where $x = 1$.

### E. A HYPOTHESIZED INSTANTIATION OF THE 'MECHANISM-PLUS-X' FRAMEWORK

Using the running example of AD research, let us develop a hypothesized manifestation of the discussed model. Based on the proposal that biological cells might have a pacemaking facility (Ehsani 2012), a speculative yet plausible 'general operative principle' might be called *synchrony* (i.e., **X = intracellular and cell-cell synchrony**). Specifically, cells might have a propensity to always move toward and facilitate a state of synchrony in their cytoplasm. The AD mechanism depicted in **Figure 2** can thus be viewed in a different light: protein aggregates might lead to a so-called phase change[39] in the cell such that portions of the liquid/viscous cytoplasm acquire a more 'solid' phase (Eguchi et al. 2018), leading areas of the cell to become *asynchronous* relative to others. The pathobiology of AD would therefore not merely be a problem of protein aggregation but rather a broader issue of cellular asynchrony. This model could be further developed theoretically, leading to new and as-yet unexplored empirical questions and potential solutions.

### F. POTENTIAL FOR PROGRESS AND LIMITATIONS

The 'mechanism-plus-X' paradigm need not be limited to one nomological factor, but could in theory be 'mechanism-plus-X-plus-Y' etc. This is a minimal way of steering clear of the pitfalls of classical mechanistic models in biology and to reintroduce sound and uncontroversial elements of non-mechanistic models back into a useful theoretical framework, perhaps a genuine move from a philosophy of the *practice of* biology, to a philosophy of biology itself and then perhaps a philosophical biology altogether.[40] This approach might additionally illuminate or demystify some of the everyday mysteriousness of biological phenomena and research. For example, in nuclear magnetic resonance (NMR) spectroscopy (a part of structural biology), there is such a thing as a "magic-angle spinning solid-state NMR". Why is the angle "magical"? The demystifying process might also sometimes involve pointing to lower-hanging fruits before approaching very big problems. For example, we read of the *Human Brain Project* that:

> "When we see a phenomenon that looks mysterious and difficult and intractable, there is a scientific possibility that what we are seeing and experiencing is a shadow projection from higher-dimensional representations […] We need mathematics to climb up into those dimensions. Then we'll understand how those shadows emerge. Consciousness may be a shadow." (Ananthaswamy 2017)

---

[38] This example is relying on a hypothesis whereby the plasma membrane of cells is proposed to have a cellular timekeeping capacity (Ehsani 2012).
[39] For the concept of 'phase separation' in cells, see (Gomes and Shorter 2018; Wang, Choi, et al. 2018).

This statement is too complex for me to comprehend, in line with the message of the earlier quotation by Frigg on mathematical models. Glymour and colleagues I think correctly remark that "one would like to know in what respects systems are sometimes too complex for people to give more than random judgments, or none at all" (Glymour et al. 2010). The framework discussed in this paper may decrease at least a little of the complexity.

What I am proposing here is only new in so far as the current mainstream practice of biology is concerned. Considering the historical developments of biology, Jessica Riskin remarks that "Biologists around the turn of the twentieth century accepted that change could take place through a combination of random and mechanically determined events, but not that change could be contingent, directed by tendencies or actions within the living organisms themselves" (Riskin 2016, 251). Therefore, the current trends can be traced back at least for more than a century. Today, a combined effort in philosophy and biology is almost exclusively reserved for analyses of various revisions of parts of evolutionary theory (e.g., (Pigliucci and Finkelman 2014), (Scholl and Pigliucci 2015), (Ramsey and De Block 2017) and (Murase and Baek 2018)). While these efforts are useful for accounts in, say, ecological models of predator-prey relationships, only the very basic and time-tested elements of evolutionary theory are understood (and capable of being utilised) at the cellular and protein/genetic level (see, e.g., (Ehsani et al. 2011) and (Halfmann and Lindquist 2010)). There are therefore many more rudimentary levels of understanding that need to be filled before reaching the level of evolutionary synthesis and similar debates.

---

[40] There are in fact analogies here to historical developments in linguistics: "The rationalist philosophy of language merged with various other independent developments in the seventeenth century, leading to the first really significant general theory of linguistic structure, namely the general point of view that came to be known as 'philosophical' or 'universal' grammar. Unfortunately, philosophical grammar is very poorly known today" (Chomsky 2006, 12-13).

There are of course limits to what this and any framework could achieve in terms of producing fruitful and workable theories of biological phenomena. One is our inevitable reliance on words borrowed from everyday language. We are what is sometimes called *Homo sapiens-L* or "humans with language" (McGilvray 2017). The semantic baggage of common linguistic phrases and our common-sense mechanistic views of the natural world together combine to impose notions such as 'contact', 'interaction', and 'neuronal firing', to name a few,[41] that may take us further away from the reality we can understand. Then there are limits to our rational, philosophising and science capacities, as Shakespeare cogently states: "There are more things in heaven and earth, Horatio, than are dreamt of in your philosophy" (*Hamlet*; Act I, scene v), and Chomsky points out the following regarding our overall mental capacities:

> "It is quite possible—overwhelmingly probable, one might guess—that we will always learn more about human life and human personality from novels than from scientific psychology. The science-forming capacity is only one facet of our mental endowment. We use it where we can but are not restricted to it, fortunately." (Chomsky 1988, 159)

Nonetheless, there is much that can be done with our rational and philosophical capacities only if one is correctly cognizant of these limits and invests one's labour where there is any conceivable chance of bearing fruit. René Descartes (1596-1650), in his unfinished 1628 treatise *Regulae ad directionem ingenii* ('Rules for the Direction of the Mind'), posits that "If in the matters to be examined we come to a step in the series of which our understanding is not sufficiently well able to have an intuitive cognition, we must stop short there. We must make no attempt to examine what follows; thus we shall spare ourselves superfluous labour" (Descartes 1997, 26). Overall, paraphrasing the subtitle of Bertrand Russell's (1872-1970) book *Human Knowledge* (Russell 1948), one should remain aware of the scope and limits of (i) the measurement instruments at hand, (ii) our own sensory systems and (iii) our cognitive and theorising capacities.

---

[41] A relevant article is: Klymkowsky M. W. "When is a gene product a protein when is it a polypeptide?" *PLOS Sci-Ed Blog*, 15 May 2018 (now available at: https://bioliteracy.blog/2018/05/15/when-is-a-gene-product-a-protein-when-is-it-a-polypeptide/).

How should we evaluate the success of our approach? This is a difficult question, because explanatory or predictive success cannot really say much about our level of understanding. A numerical model, for example, might be able to predict how a protein folds, but it teaches us little as to the actual processes involved. Russell Meyer suggests a "Woodwardian interventionism" as an assessment tool (Meyer 2018), referring to a proposal by Woodward that "causal (as opposed to merely correlational) relationships are relationships that are potentially exploitable for purposes of manipulation and control" (Woodward 2008, 219). This essentially recalls Feynman's dictum, and we can replace the word "create" with "manipulate and control" to become "what I cannot manipulate and control, I do not understand". This is potentially promising, but it depends on how the control is achieved. Again, using the example of protein folding, if we manipulate the amino acid properties heuristically (i.e., by trial and error) and achieve our set goal, I would not count that as a success in our current framework. But if the control is achieved using rational predictions from our mechanistic-nomological theories, that is indeed a success. In the meantime, however, one should remain aware of the degrees of manipulability offered by the actual circumstances at hand. For instance, the "diversity of natural microbial communities can be a major obstacle for microbiome manipulations in nature" (Weinhold et al. 2018).

Ultimately, the first unfailing measure of success in any framework is that we could simplify the problem at hand. The second measure would be using the framework to refine and expand our common-sense and intuitive notions. To take an example from mathematics and geometry, which also applies to protein shapes, the "intuitive notion of 'shape' has been made more formal and rigorous in Euclidean and other geometries".[42] And third, the central aim of our theories and models should be to move toward the unification of biology with chemistry, recalling chemistry's unification with physics around a century ago. These are equivalent to Peter Lipton's "scope, precision, mechanism, unification, and simplicity" objectives (Lipton 2007), set against the context of a general operative principle.

---

[42] Luke Fenton-Glynn, personal communication.

In Chomsky's *Aspects of the Theory of Syntax* (1965), a hierarchy of adequacies was formulated as being *observational*, *descriptive* and *explanatory* adequacies. Even at the level of explanation, which has predictive power, one could still venture outward, i.e., to elevate the *what*-properties of the phenomenon under investigation to *why* the what-properties are the way they are (Chomsky 2004). There has in fact been such a move in generative linguistics (beyond the level of explanatory theories) with the *Minimalist Program*, where Chomsky's use of the word 'program' is in reference to Imre Lakatos's (1922-1974) 'research programme' that consists of a *set of* interrelated and hierarchical theories and sub-theories. Such a programme in biology could be a natural outgrowth of the framework presented here.

Finally, an additional measure of success is if our framework would allow for some of the cognitive bounds that limit our level of understanding *to become clearer*. Given that we began by referring to Newton's undermining of mechanical philosophy, it might be apt to also end on a note about him. Chomsky writes that according to Hume, Newton's "most spectacular achievement was that while he 'seemed to draw the veil from some of the mysteries of nature, he shewed [sic] at the same time the imperfections of the mechanical philosophy; and thereby restored [Nature's] ultimate secrets to that obscurity, in which they ever did and ever will remain'" (Chomsky 2014).

## G. CONCLUSIONS

I have argued that despite the challenges to the philosophical basis of mechanistic theories, elements such as regularity/correlation and hierarchy should be preserved and incorporated into a 'mechanism-plus-X' framework, where X is an expandable nomological feature of the context of the phenomenon under study. These nomological features could take various forms, perhaps more promisingly the form of general operative principles obeying computational efficiency, cellular synchrony, etc. Moreover, we could say that biological model creation cannot be partly based on the presumed functions of the model's components; functions should only be proposed and derived *ex post facto*, as a result of the proposed 'mechanism-plus-X' framework. Such a framework could simplify our understanding of various overarching theories of the biological world and may bring us some steps closer to the realisation of a revitalised philosophical biology in line with the post-Newtonian ideals of science. The framework could make use of an array of established analytic tools in philosophy and areas of more recent active investigation, including connexive non-classical logic and mereology. This may allow for biology to again have, in the words of Carl Woese, something "fundamental to tell us about the world" (Woese 2004).

**APPENDIX**

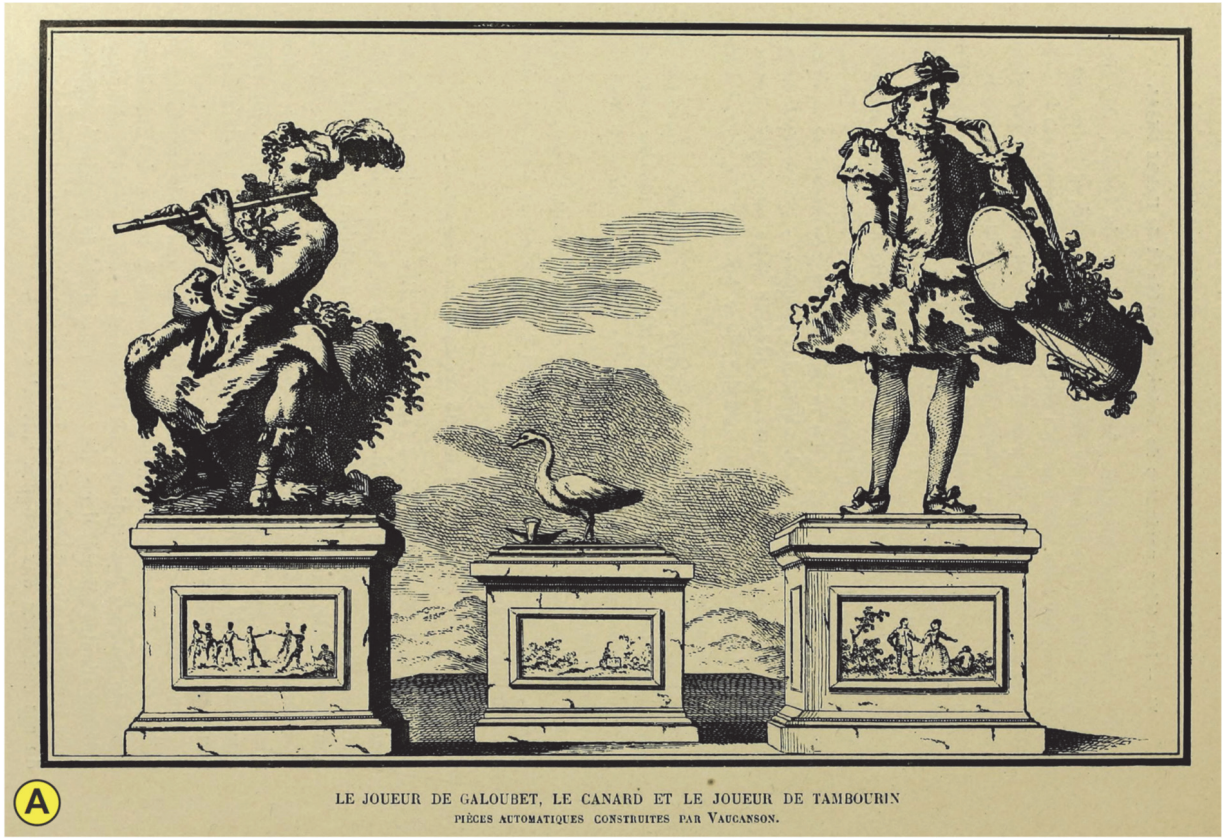

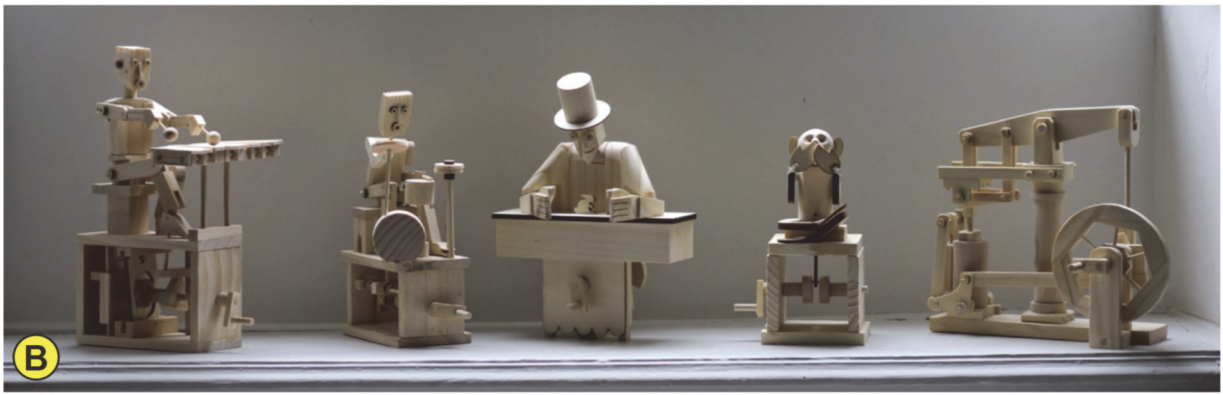

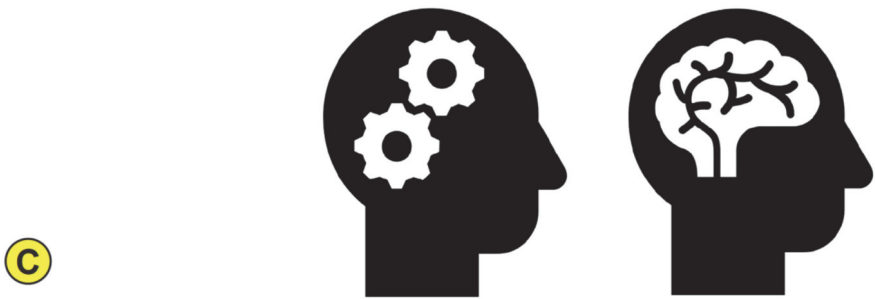

**Figure 1. Mechanistic models are linked to our intuitions about the natural world. (A)** Jacques de Vaucanson's automatic flute player, digesting duck and tambourine player remain marvels of artisanal engineering (source: *Wikimedia Commons*). **(B)** Mechanical wooden models of a xylophone player, drummer, magician, monkey and beam engine from *Timberkits* (Llanbrynmair, Wales) (source: https://www.timberkits.com/about/our-team/). The "puppeteer" could be a human hand, or an attached battery module. This is the nomological context of the models. **(C)** Today, mechanisms have become part of the iconography representing most natural faculties, including that of cognition (source: *Microsoft Word*).

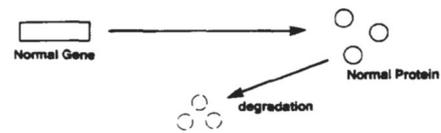
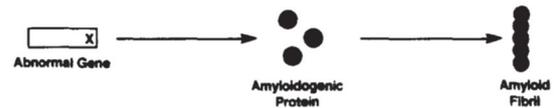
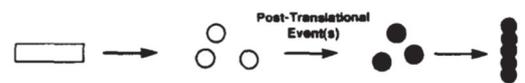
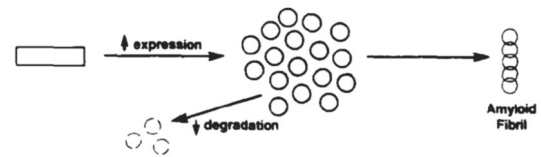
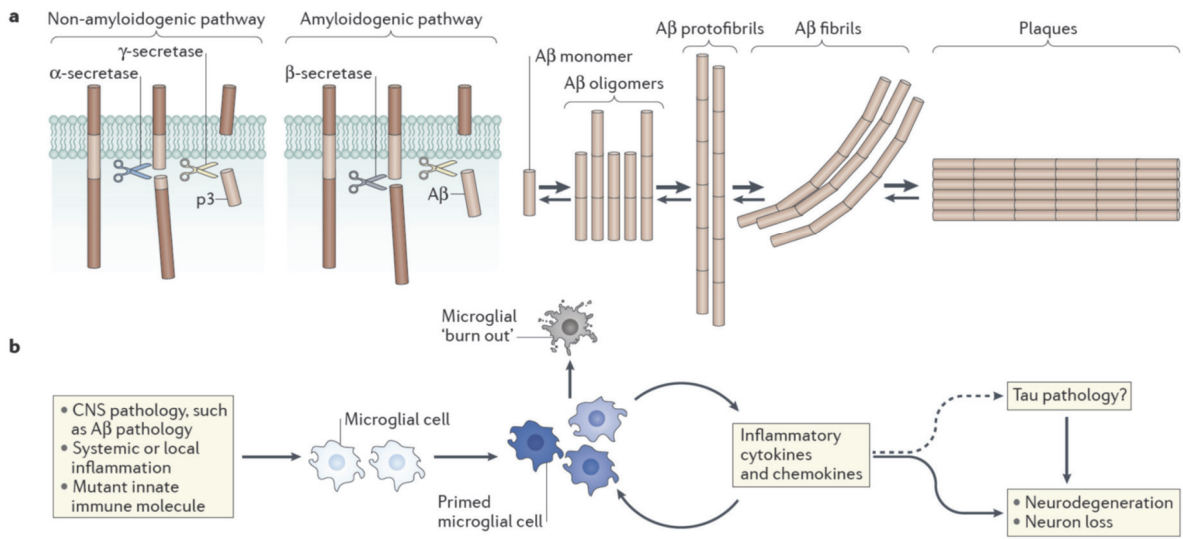

**Figure 2. Mechanism of protein aggregation in Alzheimer's disease. (A)** Original caption: "Mechanisms for amyloid fibril formation are shown schematically. Open circles represent normal protein molecules, solid circles are amyloidogenic protein molecules, and overlapping circles are the fibrils made from these proteins" (Caputo and Salama 1989). **(B)** Original caption: "The increase in production and/or reduced clearance of amyloid-β (Aβ) […] derived from the β-amyloid precursor protein" (Heppner, Ransohoff, and Becher 2015). Panel (b) here depicts possible interactions of the protein aggregates with the immune system. Reproduced with permission from *Elsevier* and *Springer Nature*, respectively.

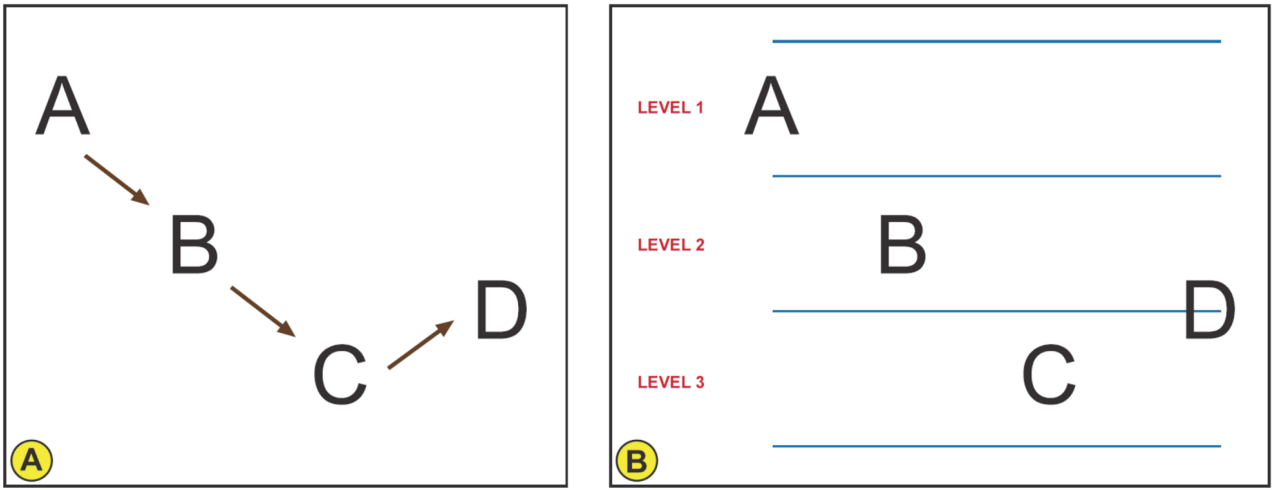

**Figure 3. Two depictions of a mechanism. (A)** A traditional depiction of a mechanistic model, where 'A' *leads* to 'B', which *leads* to 'C', which then *leads* to 'D'. This inevitably conjures a necessity of correlation and contact, where 'A' and 'B' have to interact at some close level, notwithstanding considerations as to whether or not any intermediaries (molecules or otherwise) could exist between the two. **(B)** An alternative view of thinking about the 'mechanism' in **panel A** could be to simply do away with the arrows and any notion of sequential contact. Rather, each of 'A', 'B', 'C' and 'D', as components that have arisen as *hierarchical* outputs as a result of probing a biological phenomenon, can be put into *ordered levels*, one or more of which could straddle or bridge two levels. Such straddling components could be stimulating windows of further investigation into the phenomenon under study. The levels here are spatial levels, such as intra- and extracellular spaces.